\documentclass[aps,prl,amsmath,amssymb,superscriptaddress,showpacs,twocolumn]{revtex4-1}

\usepackage{graphicx} \usepackage{color} \usepackage{verbatim} 
\usepackage{hyperref}

\newcommand{\sroinf}{SrRuO$_3$}
\newcommand{\croinf}{CaRuO$_3$}
\newcommand{\sroone}{Sr$_2$RuO$_4$}
\newcommand{\srotwo}{Sr$_3$Ru$_2$O$_7$}

\def\K{\mathrm{K}}
\def\eV{\mathrm{eV}}

\begin{document}

\title{ Transport properties of Metallic Ruthenates: a DFT+DMFT
  investigation }

\author{Xiaoyu Deng}
\affiliation{Department of Physics and Astronomy, Rutgers University,
  Piscataway, New Jersey 08854, USA}
\author{Kristjan Haule}
\affiliation{Department of Physics and Astronomy, Rutgers University,
  Piscataway, New Jersey 08854, USA}

\author{Gabriel Kotliar}
\affiliation{Department of Physics and Astronomy, Rutgers University,
  Piscataway, New Jersey 08854, USA}

\date{\today}

\begin{abstract}
  We present a systematical theoretical study on the transport
  properties of an archetypal family of Hund's metals, \sroone{},
  \srotwo{}, \sroinf{} and \croinf{}, within the combination of first
  principles density functional theory and dynamical mean field
  theory. The agreement between theory and experiments for optical
  conductivity and resistivity is good, which indicates that
  electron-electron scattering dominates the transport of
  ruthenates. We demonstrate that in the single-site dynamical mean
  field approach the transport properties of Hund's metals fall into
  the scenario of "resilient quasiparticles". We explains why the
  single layered compound \sroone{} has a relative weak correlation
  with respect to its siblings, which corroborates its good
  metallicity.

\end{abstract}
\pacs{71.27.+a, 72.10.-d, 78.20.-e}

\maketitle 

The anomalous transport properties in correlated metals have been of
great interest for many decades, but the main focus was the
electron-electron scattering due to Hubbard-like short range
repulsion. Many features found in realistic materials are well
captured by the dynamical mean field theory (DMFT) in simple Hubbard
models\cite{georges1996,pruschke1995}, for example, the very low
coherence scale $T_{FL}$ below which Fermi liquid (FL) theory holds
\cite{nozieres1997}, and the high temperature "bad metal" behavior
that resistivity exceeds the Mott-Ioffe-Regel limit\cite{emery1995}.
In the broad crossover regime between FL and "bad metal" the
"resilient quasiparticles" which survive up to high
temperature\cite{deng2013, xu2013} dominate the transport. Very
recently Hund's metals\cite{haule2009-c} have attracted a lot of
attentions. These are materials in which the Hund's interaction rather
than the Hubbard repulsion gives rise to the heavy quasiparticle mass
in several transition metal compounds such as iron
pnictides\cite{yin2011-a} and ruthenates\cite{mravlje2011,
  yin2012}. The understanding of the scattering mechanism in Hund's
metals and their consequences for the transport properties have not
been explored much.

In this paper we study the archetypal Hund's metals\cite{mravlje2011,
  yin2012, georges2013}, the metallic members of Ruddlesden-Popper
serials of ruthenates (A$_{n+1}$B$_{n}$O$_{3n+1}$): \sroone{} ($n=1$),
\srotwo{} ($n=2$), \sroinf{} and \croinf{} ($n=\infty$).  Ruthenates
have been extensively studied as prototypical strongly correlated
systems, with large effective mass enhancements revealed by various
experiments\cite{allen1996,cao1997, shepard1997, maeno1997, ikeda2000,
  bergemann2003, alexander2005, tamai2008, mercure2009, iwasawa2012,
  shai2013, veenstra2013, allan2013, schneider2014}.  They exhibit a
very small coherence scale $T_{FL}$, as well as a crossover into "bad
metal" regime\cite{cao2004,hussey1998, tyler1998,schneider2014,
  bruin2013,klein1996}. Surprisingly the single layered compounds
\sroone{} is more metallic than the pseudocubic \sroinf{} and
\croinf{} at relative low temperature ($\leq 400\K $). This is
different from many other systems, for example, the Ruddlesden-Popper
family of strontium vanadates, lanthanum nickelates, lanthanum
cuprates, strontium iridates, where the single layered compounds are
insulating and the pseudocubic ones are metallic.

The method used in this paper is the combination of density functional
theory and DMFT (DFT+DMFT) in the charge self-consistent and all
electron formulation that avoids building the low energy Hubbard
model, which is successful in the quantitative descriptions of
electronic structures in many correlated systems\cite{kotliar2006}.
There are a few DFT+DMFT studies on ruthenates available in the
literature\cite{jakobi2011, mravlje2011, dang2014, dang2015} but they
are performed on low energy Hubbard models. Moreover a complete
investigation of the transport properties within a uniform DFT+DMFT
scheme for these Hund's metals is missing.  We show that the transport
properties of these materials, especially their temperature
dependence, are related to the underlying "resilient
quasiparticles". We find that \sroone{} is the least correlated one
among the compounds considered, which corroborates its good
metallicity.

We carry out the DFT+DMFT calculations with the all-electron DMFT as
implemented in Ref.~{\cite{haule2010}} based on Wien2K
package\cite{blaha2001}. The continuous-time quantum Monte-Carlo
method with hybridization expansion is used to solve the impurity
problem\cite{haule2007,werner2006}. A large energy window 20$\eV$ is
used to construct the atomic-like localized $d$ orbitals. This
procedure permits us to use the same interaction parameters for all
the ruthenates. We estimate the Slater integral within the localized
orbitals assuming a screened Yukawa-form Coulomb potential
$e^{-r/\lambda}/r$. With a proper choice of $\lambda$ we have
$(F^0,F^2,F^4)=(4.5, 8.0, 6.5)\eV$, which amounts to
$(U,J)=(4.5,1.0)\eV$. We note that in previous
studies\cite{jakobi2011, mravlje2011, dang2014} smaller interaction
parameters are used, because there the local orbitals are constructed
in much smaller energy windows thus more extended. As shown below for
all the compounds our current choice gives results in agreement with
experiments and similar quasiparticle mass enhancements as in previous
studies. The standard double counting in the fully localized limit
form is adopted. The resistivity and optical conductivity are
calculated using formalism of Ref.\cite{haule2010} in which the
vertext corrections to the transport are neglected. Both polynomial
fitting to the low frequency data and maximum entropy method are used
to analytically continue the computed self energy. 
We focus on the paramagnetic states only and neglect the
ferromagnetism in \sroinf{} at low temperature.

We first justify our choice of interaction parameters by examining the
effective mass enhancement, which is computed by
$m^*_\mathrm{theory}/m_\mathrm{DFT}=1/Z=1-\frac{\partial
  \mathrm{Re}{\Sigma(\omega)}}{\partial\omega}|_{\omega= 0}$. These
values extracted at $T=58\K$ are presented in Table.\ref{FL_para},
along with their experimental estimations. For \croinf{} and
\sroone{}, they are in good agreement with experiments, and with
previous DFT+DMFT calculations\cite{jakobi2011, mravlje2011,
  dang2014}. For \sroinf{} no comparison is available since
measurements are performed in the ferromagnetic state. Our result
shows that the correlation strength of \sroinf{} is close to the one
of \croinf{} despite that the latter has a larger distortion and
slightly narrower bandwidth.  The correlation is stronger in the
considered paramagnetic phase than that in the experimental
ferromagnetic phase, which is generally expected since magnetism tends
to reduce correlation. \srotwo{} is peculiar where a strong
momentum-dependence of the effective mass enhancement is revealed by
high quality quantum oscillation (QO) and angular-resolved
photoemission (ARPES) measurements \cite{tamai2008,allan2013}. The
strong momentum dependence is beyond our single-site DMFT
approach. However our calculation gives a value very close to the mass
enhancement ($\sim 6$) on a large portion of the Fermi surface as
found by ARPES \cite{allan2013}. Since the theoretical mass
enhancements across all the materials agree reasonably well with
available experimental values, the current choice of parameters is
satisfactory. We note that \sroone{} has smaller effective mass
enhancements that its siblings, in agreement with experiments.

%%%% Table
\begin{table}[!hbt]
\begin{center}
%\begin{ruledtabular}
\squeezetable
\begin{tabular}{c c c c c c c c }
  \toprule
  &  Sr$_2$RuO$_4$     & Sr$_3$Ru$_2$O$_7$  & SrRuO$_3$              & CaRuO$_3$          \\ \hline
%  Order              &  PM                &   PM               & FM ($T<160K$)              & PM                 \\ \hline
%  $T_{FL}$            &  $\simeq 25\K$     & $\simeq 10\K$      & $\simeq 15\K$
%  &$\simeq 1.5\K$\cite{schneider2014}           \\ \hline
  $m^*_\mathrm{theory}/m_\mathrm{DFT}$      &  4.3 ($xz/yz$)
  & 6.3 ($xz/yz$)      & 6.6     &  6.7         \\  
  &   5.6 ($xy$)  & 6.4 ($xy$)      &      &           \\  \hline
  $\gamma_\mathrm{exp}/\gamma_\mathrm{DFT}$    & 4                & 9                & 3.7 (FM)               &  6.5             \\ \hline
  $m^*_{\mathrm{ARPES}}/m_\mathrm{DFT}$    &   $\simeq 3 $\cite{iwasawa2012,veenstra2013}
  & $\simeq 6 $ \cite{allan2013}  &                                \\ \hline
  $m^*_\mathrm{QO}/m_\mathrm{DFT}$     &  3, 3.5 ($xz/yz$)  &
  & & 6.1 \cite{schneider2014}                  \\ 
  &   5.5 ($xy$)\cite{bergemann2003}  &
  \\ \hline
  \botrule
\end{tabular}
%\end{ruledtabular}
\end{center}
\caption{The mass enhancement of ruthenates obtained in current DFT+DMFT
  calculations at $T=58\K$.
  Values estimated from specific heat coefficients, ARPES and
  quantum-oscillation measurements are presented for comparison. The experimental specific heat
  coefficients $\gamma_\mathrm{exp}$ are
  taken from Ref \onlinecite{allen1996,cao1997,
    shepard1997, maeno1997, ikeda2000}, while the
  corresponding DFT values are computed with
  Wien2K. $m^*_\mathrm{theory}/m_\mathrm{DFT}$ of \sroinf{} and \croinf{} is averaged over
  $t_{2g}$ orbitals . We note the
  $m^*_{\mathrm{ARPES}}/m_\mathrm{DFT}$ of \srotwo{} is taken from a
  large fraction of the Fermi surface as discussed in the text. The $m^*_\mathrm{QO}/m_\mathrm{DFT}$ of \croinf{} is the value at
  zero magnetic field estimated from the data  in
  Ref.~\onlinecite{schneider2014} assuming Kadowaki-Woods relation.}
\label{FL_para}
\end{table}

\begin{figure}
\includegraphics[width=0.9\columnwidth]{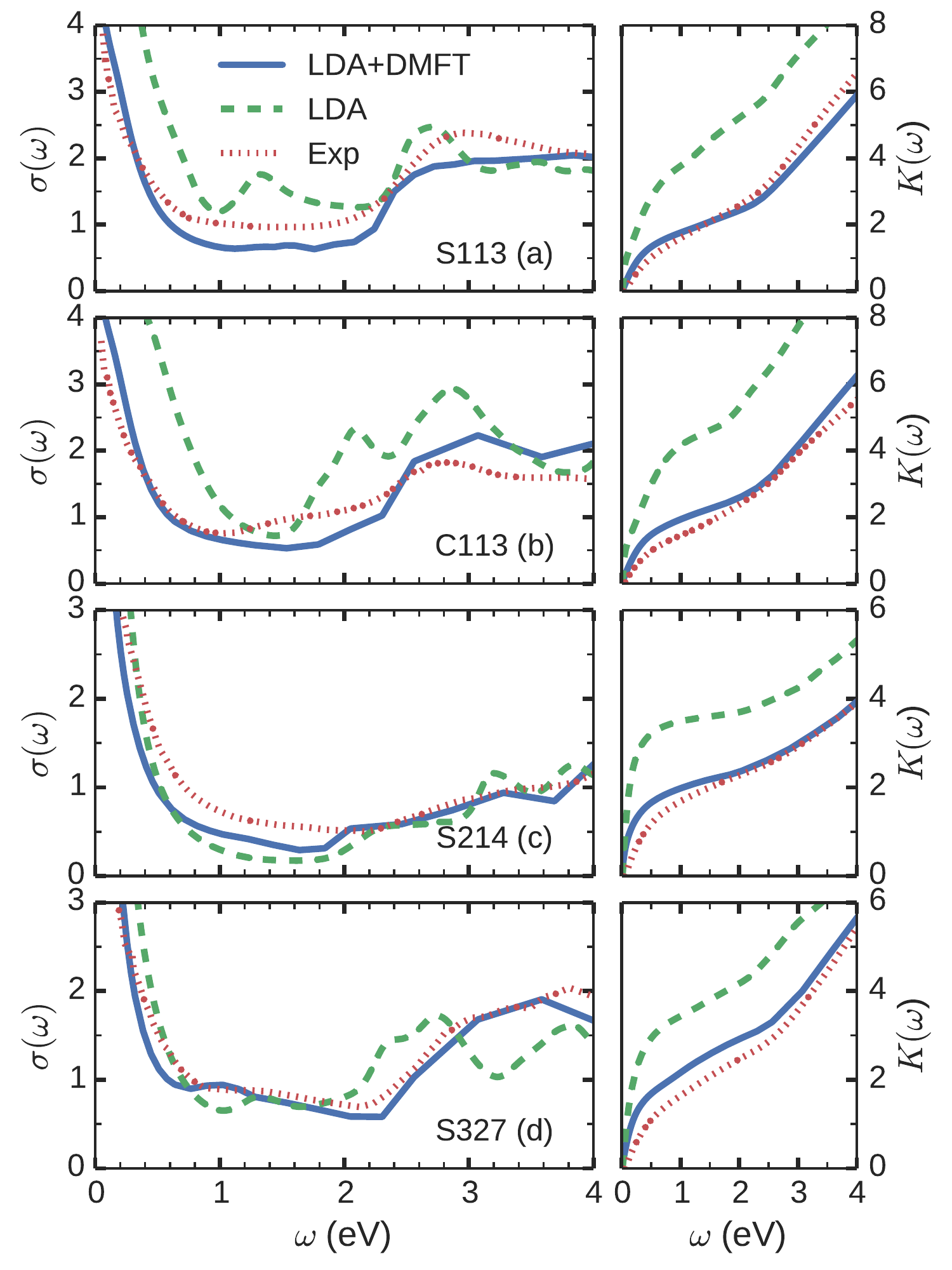}
\caption{The optical conductivity (left panel) and the corresponding
  integrated spectral weight (right panel) of ruthenates calculated
  within DFT+DMFT method ($T=298\K$) and DFT. Experimental data at
  room temperatur are taken from \onlinecite{lee2004-a} for
  comparison. S113, C113, S214, S327 are acronyms for \sroinf{},
  \croinf{}, \sroone{} and \srotwo{}. }
\label{Optics} 
\end{figure}

The computed optical conductivities at room temperature are shown in
Fig.~\ref{Optics} along with the experimental measurements and DFT
results.  Our calculated optical conductivities are consistent with
the experiment measurements for all the compounds considered, and DMFT
improves systematically the DFT results. The height and width of the
Drude response are reasonably captured in our calculations. We note
that the Drude response contains not only intra-orbital but also
inter-orbital transition among $t_{2g}$ orbitals, which is argued to
be important for the $\omega^{-1/2}$ behavior in \croinf{}
\cite{dang2014}. A broad peak centered around $3\eV$ appears in all
the compounds as observed in experiments. The broad peak is assigned
to the transition between the O-$2p$ to Ru-$d$ orbitals. Note that DFT
predicts an additional peak in \sroinf{} at about $1.5\eV$ and in
\croinf{} at about $2.0\eV$, which can be assigned to $t_{2g}$-$e_g$
transition. The amplitude of $t_{2g}$-$e_{g}$ transition depends on
the extent of GdFeO$_3$ distortion, and is insignificant or even
missing in \srotwo{} and \sroone{}, likely due to the matrix-element
effects. However these peaks are shifted to higher frequency and
merged with the broad peak at $3\eV$ in our DFT+DMFT calculations in
agreement with experiments.

The spectral weight distributions, computed as $K(\omega)=\int_0^{\omega}\sigma(\omega')d\omega'$,
from both experimental and calculated optical conductivities, are also
depicted in Fig.~\ref{Optics}. Strong correlations normally induce an
anomalous spectral weight redistribution, which is the case in
ruthenates. Compared with the DFT spectral weight distribution, a
significant reduction is seen in the experimental data for all the
ruthenates. A fraction of spectral weight is transferred to much
higher frequency ($\geq 4\eV$). The current DFT+DMFT calculations give
spectral weight distributions in good agreements with experiments and
capture the spectral weight reduction of LDA band theory nicely for
all the compounds. The good agreements between theory and experiments
of both optical conductivity and spectral weight distribution are
solid evidences that electron-electron correlations dominate the
electron dynamics in ruthenates.

\begin{figure}
\includegraphics[width=0.7\columnwidth]{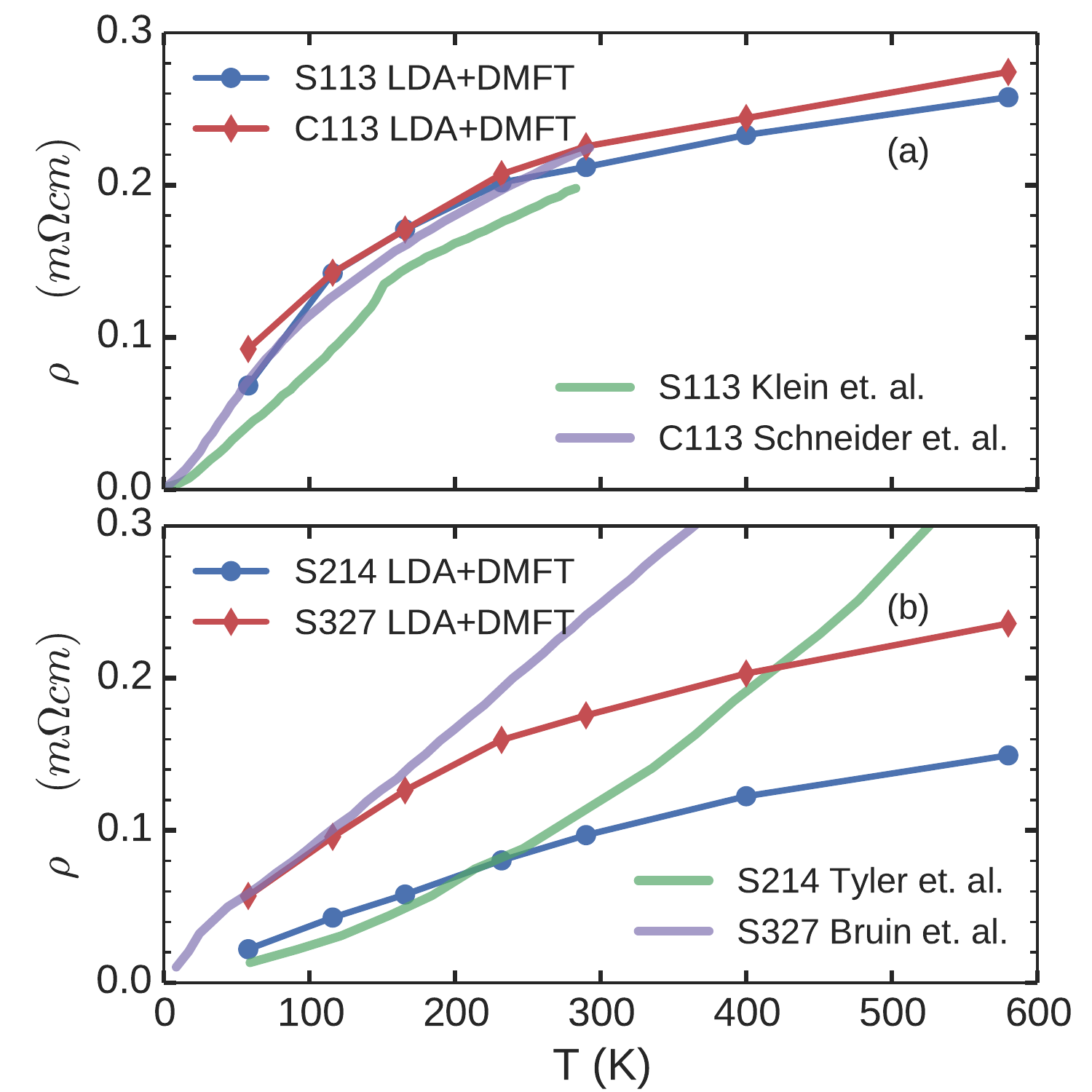}
\caption{The calculated resistivity of ruthenates with DFT+DMFT method
  for different temperature. The experimental measurements shown for
  for comparison are taken from
  \onlinecite{klein1996,schneider2014,tyler1998,bruin2013}. }
\label{Res}
\end{figure}

Now we focus on the resistivity of these compounds which is directly
related to zero-frequency limit of the optical response. The results
are depicted in Fig.\ref{Res} and compared with experiments.  We note
that in our calculations the resistivity of \sroinf{} (\croinf{}) has
a relative small anisotropy (less than $15\%$), in accordance with
experimental determinations\cite{genish2007,proffit2008}, therefore
only its average over three principle axis is presented. For \croinf{}
the agreement between the calculated and measured resistivity is
almost perfect in both the overall scale and the temperature
dependence in the whole temperature range. The shoulder at around
$200\K$ which marks the substantial change of the slope of the
resistivity is well captured. For \sroinf{} the calculated resistivity
is very close to the one of \croinf{}. Its agreement with experiment
is very good above the Curie temperature $T_c\sim 160\K$, however below $T_c$
there is extra reduction of resistivity due to restoration of
coherence in ferromagnetic state which is neglected in our
calculations.

The agreement between the computed in-plane resistivities of the
layered compounds \sroone{} and \srotwo{} and the experiments as shown
in Fig.~\ref{Res}(b), is not as good as for \croinf{}. The calculated
resistivities have similar temperature dependence as those of the
pseudocubic compounds with a shoulder at around $200\sim
300\K$. However the measured ones are different. The resistivity of
\srotwo{} is almost linear in temperature up to $300\K$ with a weak
shoulder showing up at low temperature (around $20\K$), and that of
\sroone{} does not exhibit a shoulder at all. Nevertheless there are
three features correctly captured in our calculations. The resistivity
of both compounds agree reasonably in the overall scale with
experiments, especially at relative low temperature. The resistivity
shows no sign of saturation at high temperature, although the
increasing is not as fast as found in experiments. And going from
pseudocubic structure to layered structure, the material becomes more
conductive.

Despite the difference in the coherence scale, the computed
resistivity of ruthenates where Hund's coupling dominates the
correlations, has a very similar shape to the one of single band doped
Hubbard model where Hubbard repulsion dominates the
correlation\cite{deng2013}. Therefore this anomalous shape is likely a
characteristic of the resistivity in single-site DMFT approach when
the vertex corrections to the transport are neglected. Our results
show that the vertex corrections are small in the pseudocubic
ruthenate, but they are likely the cause of the descrepancy between
this theory and experiments in layered compounds. Other effects such
as electronc-phonon scattering and nonlocal interactions might also
play some role.

We note that in agreement with experiments, both \sroone{} and
\srotwo{} exhibit strong anisotropy in our LDA+DMFT calculations that
the calculated out-of-plane resistivity is orders of magnitude larger
than the in-plane one. The large anisotropy comes from the anisotropy
of the plasma frequency which is captured by DFT\cite{singh1995} and
also presents in DFT+DMFT.
\begin{figure}
\includegraphics[width=0.7\columnwidth]{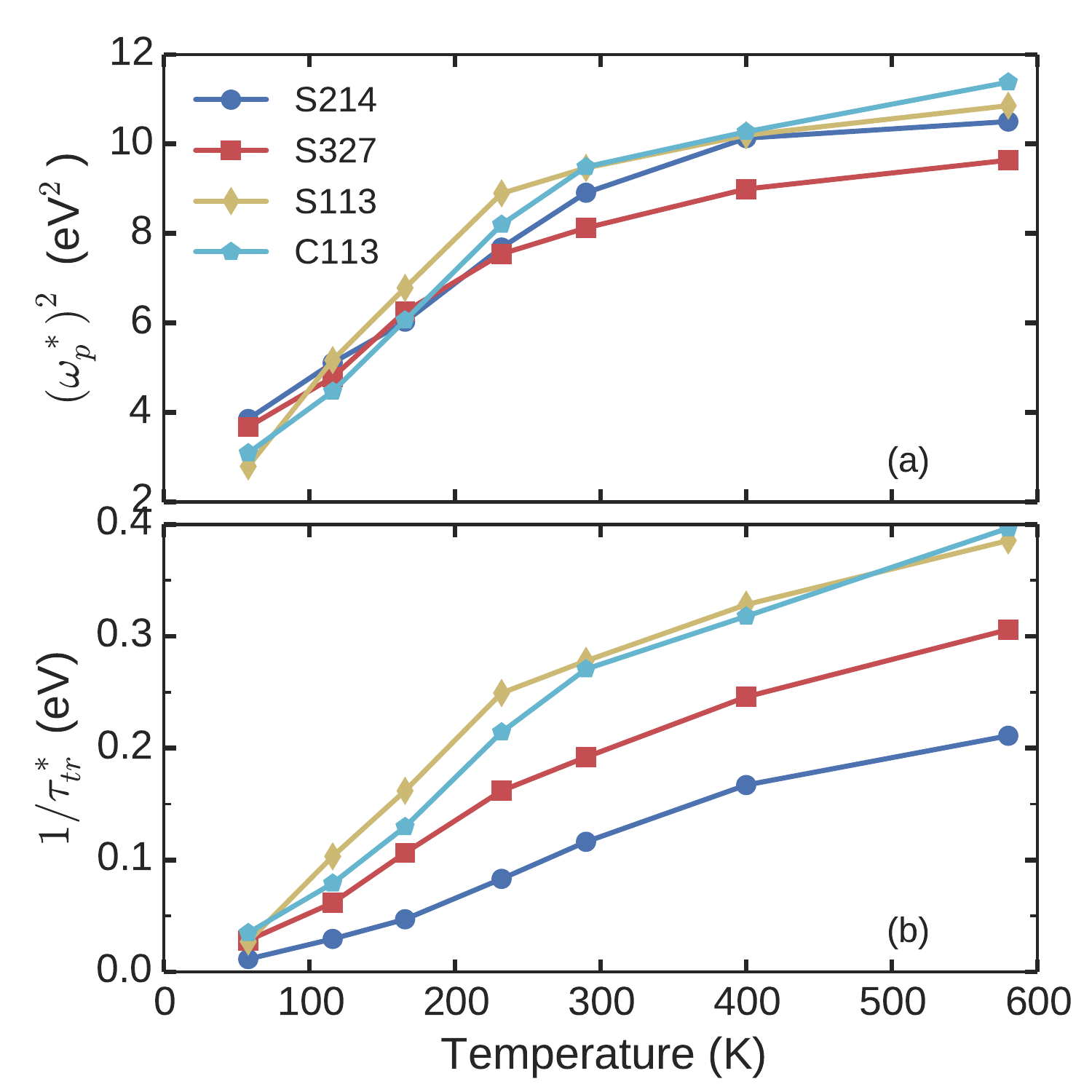}
\caption{The effective plasma freqeuency square $(\omega_p^*)^2$ and the
  effective quasiparticle scattering rate $1/\tau_{tr}^*$ extracted
  from the computed optical conductivity with DFT+DMFT method
  according to the formalism in Ref.~\onlinecite{deng2014}.}
\label{OpticDrude}
\end{figure}

\begin{figure}
\includegraphics[width=0.7\columnwidth]{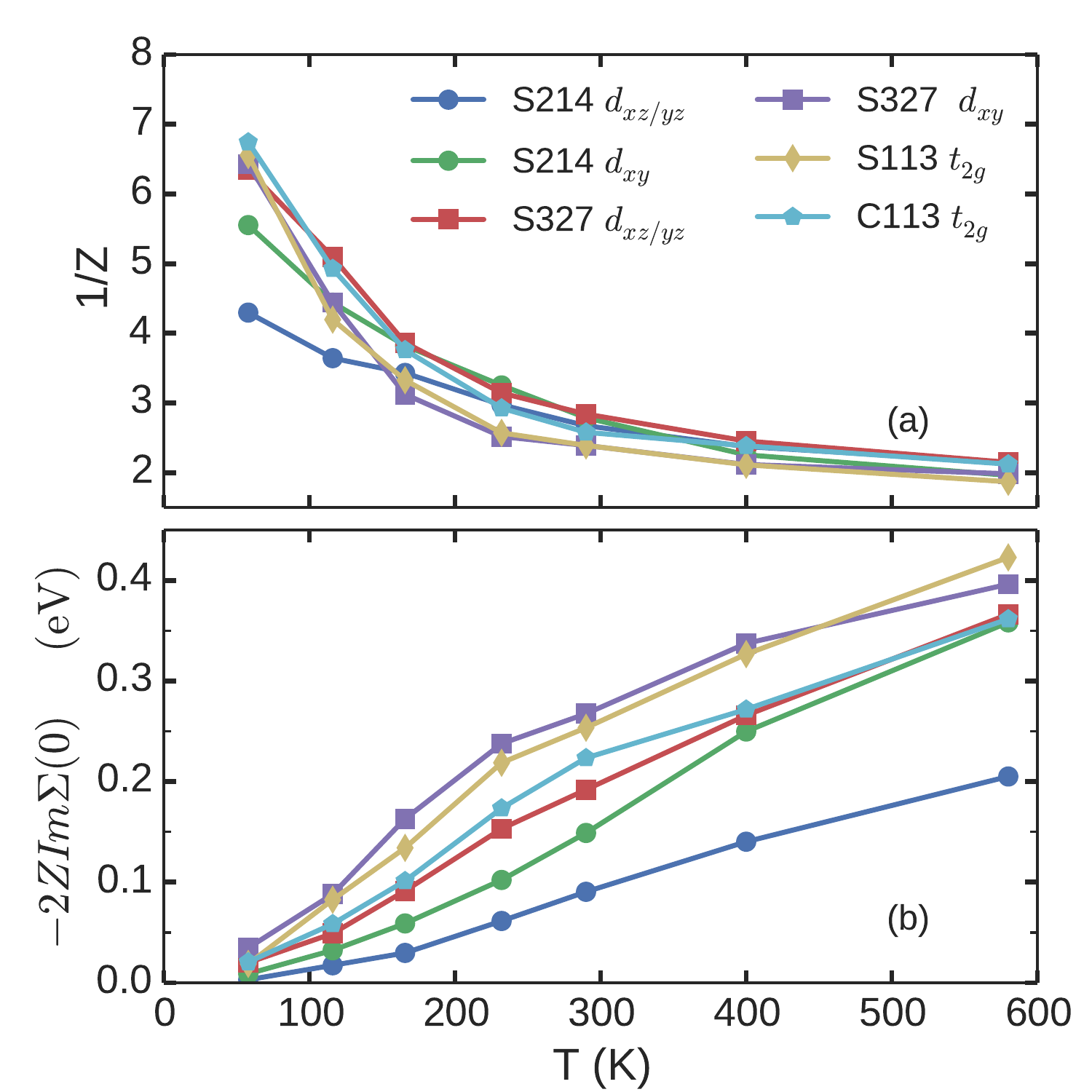}
\caption{The calculated effective mass enhancement
  $m^*_\mathrm{theory}/m_\mathrm{DFT}=1/Z$ and the effective
  quasiparticle scattering rate $\Gamma^*=-2ZIm\Sigma(0)$ of different
  orbitals in ruthenates. }
\label{QP}
\end{figure}

The relatively good metallicity of the layered compounds \sroone{}
with respect to its siblings is captured in our calculations. To gain
more understanding we recall that the $dc$ conductivity can be written
as $\sigma=(\omega_p^*)^2\tau_{tr}^*/4\pi$, where the effective plasma
frequency $\omega_p^*$ and the effective scattering rate
$1/\tau_{tr}^*$ can be extracted from the computed optical
conductivity \cite{deng2014}.  As shown in Fig.~\ref{OpticDrude},
there is strong temperature dependence of $\omega_p^*$ and
$1/\tau_{tr}^*$ in all the compounds, which are characteristics of
underlying "resilient quasiparticles"\cite{deng2014}. Interesting
unlike that of V$_2$O$_3$ in our previous study, $(\omega_p^*)^2$ in
ruthenates shows a saturation (or weak temperature dependence) above
$T\simeq 200\K$. This is possibly a characteristic of Hund's metal and
needs to be justified in further studies. As discussed in
Ref.~\onlinecite{deng2014}, $(\omega_p^*)^2$ and $1/\tau_{tr}^*$ are
directly related to $1/Z$ and the quasiparticle scattering rate
$\Gamma^*= -2Z\mathrm{Im} \Sigma(0)$ , which have also strong
temperature dependences as shown in Fig.\ref{QP} for different
orbitals. $1/Z$'s decreases when the temperature increases as found in
previous study\cite{deng2013,xu2013,deng2014}, and interestingly all
approach approximately $2$ at high temperature. The temperature
dependence of $1/Z$ is consistent with that of effective optical mass
inferred from THz conductivity of \croinf{}\cite{kamal2005}.  In
addition, we note both $1/\tau_{tr}^*$ and $\Gamma^*$ generally show
hidden Fermi liquid behavior at relative low temperature that they are
approximately parabolic in temperature \cite{xu2013,deng2014},
although the behavior is elusive in \sroinf{}.

\sroone{} is the least correlated one in the ruthenates family
according to the relative order of $1/\tau_{tr}^*$. To understand the
relative correlation strength in ruthenates, we look into the
orbital-resolved quantities, the low temperature effective mass
enhancement $m^*_\mathrm{theory}/m_\mathrm{DFT}$ in
Table.\ref{FL_para} and the quasiparticle scattering rate $\Gamma^*$
in Fig.~\ref{QP}(b). We find that the $d_{xz/yz}$ orbitals in
\sroone{} are the special ones with significantly smaller
$m^*_\mathrm{theory}/m_\mathrm{DFT}$ and $\Gamma^*$ than the others
. The uniqueness of $d_{xz/yz}$ orbitals in \sroone{} can be traced
back to their one-dimensional nature. Due to quantum confinement by
Sr-O double-layer along out-of-plane axis, these orbitals have 1D
singularities at their band edges, and a low density of states near
the Fermi level with respect to the other orbitals. The relative weak
correlation strength in these orbitals can be understood within the
same argument of Ref.\onlinecite{mravlje2011}, that the lower density
of states $\rho_{F}$ near the Fermi level implies stronger Weiss
function in DMFT, $\mathrm{Im}\Delta(\omega\rightarrow 0)\simeq
-\frac{1}{\pi\rho_F}$, and results in weaker correlation. We note that
this argument holds because in ruthenates the real part of the local
Green's functions $\mathrm{Re}G_{loc}(\omega)$ are much smaller than
the imaginary part $\mathrm{Im}G_{loc}(\omega)=-\pi\rho_{F}$ near
Fermi level\cite{supp}. As $n$ increases from \sroone{} ($n=1$), the
density of states of $d_{xz/yz}$ orbitals near the Fermi level
increases due to the relaxation of quantum confinement and the
rotation of oxygen octahedra, therefore the correlation is enhanced.
Meanwhile orbital differentiation is reduced so that eventually the
$d_{xz/yz}$ orbitals become nearly degenerate with $d_{xy}$ orbital in
pseudocubic compounds thus exhibit similar correlations. However
considering only $d_{xy}$ orbitals (as well as $d_{xz/yz}$ orbitals in
pseudocubic compounds due to the nearly degeneracy), we find that
their Weiss functions do not correlate with their relative correlation
strength. Rather the effective mass enhancement in these orbitals is
mostly related to the in-plane Ru-O bond length and the rotation of
oxygen octahedra\cite{supp}. The $d_{xy}$ orbital in \sroone{} is
slightly less correlated than the others because of the short in-plane
Ru-O bond length and the absence of oxygen octahedron rotations in
this compound. Our findings may shed light on the correlation effects
in ruthenate thin films and heterostructures where the quantum
confinement\cite{chang2009}, the Ru-O bond length and the distortions
of oxygen octahedra could be engineered.

In conclusion, our DFT+DMFT calculations provide a quite accurate
description of the transport properties in ruthenates. We demonstrate
that the resilient quasiparticle scenario is valid beyond Hubbard-like
repulsion in particular in Hund's metals. We explain the origin of the
relative good metallicity in \sroone{}. Our results also suggests that
effects such as vertex corrections, electron-phonon interactions or
nonlocal interactions would need to be considered for more precise
predictions of the resistivity of layered ruthenates.

We thank A. Georges and J. Mravlje for very useful discussions.  We
acknowledge supports by NSF DMR-1308141 (X. D. and G.K), NSF DMR
1405303 (K. H.).

{\it Note:} When preparing the manuscript we are aware of
Ref.\cite{dang2015} , which presents similar results on the
temperature dependence of the effective mass enhancements in \sroinf{}
and \croinf{} as in our study.


\begin{thebibliography}{46}%
\makeatletter
\providecommand \@ifxundefined [1]{%
 \@ifx{#1\undefined}
}%
\providecommand \@ifnum [1]{%
 \ifnum #1\expandafter \@firstoftwo
 \else \expandafter \@secondoftwo
 \fi
}%
\providecommand \@ifx [1]{%
 \ifx #1\expandafter \@firstoftwo
 \else \expandafter \@secondoftwo
 \fi
}%
\providecommand \natexlab [1]{#1}%
\providecommand \enquote  [1]{``#1''}%
\providecommand \bibnamefont  [1]{#1}%
\providecommand \bibfnamefont [1]{#1}%
\providecommand \citenamefont [1]{#1}%
\providecommand \href@noop [0]{\@secondoftwo}%
\providecommand \href [0]{\begingroup \@sanitize@url \@href}%
\providecommand \@href[1]{\@@startlink{#1}\@@href}%
\providecommand \@@href[1]{\endgroup#1\@@endlink}%
\providecommand \@sanitize@url [0]{\catcode `\\12\catcode `\$12\catcode
  `\&12\catcode `\#12\catcode `\^12\catcode `\_12\catcode `\%12\relax}%
\providecommand \@@startlink[1]{}%
\providecommand \@@endlink[0]{}%
\providecommand \url  [0]{\begingroup\@sanitize@url \@url }%
\providecommand \@url [1]{\endgroup\@href {#1}{\urlprefix }}%
\providecommand \urlprefix  [0]{URL }%
\providecommand \Eprint [0]{\href }%
\providecommand \doibase [0]{http://dx.doi.org/}%
\providecommand \selectlanguage [0]{\@gobble}%
\providecommand \bibinfo  [0]{\@secondoftwo}%
\providecommand \bibfield  [0]{\@secondoftwo}%
\providecommand \translation [1]{[#1]}%
\providecommand \BibitemOpen [0]{}%
\providecommand \bibitemStop [0]{}%
\providecommand \bibitemNoStop [0]{.\EOS\space}%
\providecommand \EOS [0]{\spacefactor3000\relax}%
\providecommand \BibitemShut  [1]{\csname bibitem#1\endcsname}%
\let\auto@bib@innerbib\@empty
%</preamble>
\bibitem [{\citenamefont {Georges}\ \emph {et~al.}(1996)\citenamefont
  {Georges}, \citenamefont {Kotliar}, \citenamefont {Krauth},\ and\
  \citenamefont {Rozenberg}}]{georges1996}%
  \BibitemOpen
  \bibfield  {author} {\bibinfo {author} {\bibfnamefont {A.}~\bibnamefont
  {Georges}}, \bibinfo {author} {\bibfnamefont {G.}~\bibnamefont {Kotliar}},
  \bibinfo {author} {\bibfnamefont {W.}~\bibnamefont {Krauth}}, \ and\ \bibinfo
  {author} {\bibfnamefont {M.~J.}\ \bibnamefont {Rozenberg}},\ }\href {\doibase
  10.1103/RevModPhys.68.13} {\bibfield  {journal} {\bibinfo  {journal} {Reviews
  of Modern Physics}\ }\textbf {\bibinfo {volume} {68}},\ \bibinfo {pages} {13}
  (\bibinfo {year} {1996})}\BibitemShut {NoStop}%
\bibitem [{\citenamefont {Pruschke}\ \emph {et~al.}(1995)\citenamefont
  {Pruschke}, \citenamefont {Jarrell},\ and\ \citenamefont
  {Freericks}}]{pruschke1995}%
  \BibitemOpen
  \bibfield  {author} {\bibinfo {author} {\bibfnamefont {T.}~\bibnamefont
  {Pruschke}}, \bibinfo {author} {\bibfnamefont {M.}~\bibnamefont {Jarrell}}, \
  and\ \bibinfo {author} {\bibfnamefont {J.~K.}\ \bibnamefont {Freericks}},\
  }\href {\doibase 10.1080/00018739500101526} {\bibfield  {journal} {\bibinfo
  {journal} {Adv. Phys.}\ }\textbf {\bibinfo {volume} {44}},\ \bibinfo {pages}
  {187} (\bibinfo {year} {1995})}\BibitemShut {NoStop}%
\bibitem [{\citenamefont {Nozi\'{e}res}(1997)}]{nozieres1997}%
  \BibitemOpen
  \bibfield  {author} {\bibinfo {author} {\bibfnamefont {P.}~\bibnamefont
  {Nozi\'{e}res}},\ }\href@noop {} {\emph {\bibinfo {title} {Theory Of
  Interacting Fermi Systems}}}\ (\bibinfo  {publisher} {{Addison-Wesley},
  Reading, {MA}},\ \bibinfo {year} {1997})\BibitemShut {NoStop}%
\bibitem [{\citenamefont {Emery}\ and\ \citenamefont
  {Kivelson}(1995)}]{emery1995}%
  \BibitemOpen
  \bibfield  {author} {\bibinfo {author} {\bibfnamefont {V.~J.}\ \bibnamefont
  {Emery}}\ and\ \bibinfo {author} {\bibfnamefont {S.~A.}\ \bibnamefont
  {Kivelson}},\ }\href {\doibase 10.1103/PhysRevLett.74.3253} {\bibfield
  {journal} {\bibinfo  {journal} {Physical Review Letters}\ }\textbf {\bibinfo
  {volume} {74}},\ \bibinfo {pages} {3253} (\bibinfo {year}
  {1995})}\BibitemShut {NoStop}%
\bibitem [{\citenamefont {Deng}\ \emph {et~al.}(2013)\citenamefont {Deng},
  \citenamefont {Mravlje}, \citenamefont {\v{Z}itko}, \citenamefont {Ferrero},
  \citenamefont {Kotliar},\ and\ \citenamefont {Georges}}]{deng2013}%
  \BibitemOpen
  \bibfield  {author} {\bibinfo {author} {\bibfnamefont {X.}~\bibnamefont
  {Deng}}, \bibinfo {author} {\bibfnamefont {J.}~\bibnamefont {Mravlje}},
  \bibinfo {author} {\bibfnamefont {R.}~\bibnamefont {\v{Z}itko}}, \bibinfo
  {author} {\bibfnamefont {M.}~\bibnamefont {Ferrero}}, \bibinfo {author}
  {\bibfnamefont {G.}~\bibnamefont {Kotliar}}, \ and\ \bibinfo {author}
  {\bibfnamefont {A.}~\bibnamefont {Georges}},\ }\href {\doibase
  10.1103/PhysRevLett.110.086401} {\bibfield  {journal} {\bibinfo  {journal}
  {Physical Review Letters}\ }\textbf {\bibinfo {volume} {110}},\ \bibinfo
  {pages} {086401} (\bibinfo {year} {2013})}\BibitemShut {NoStop}%
\bibitem [{\citenamefont {Xu}\ \emph {et~al.}(2013)\citenamefont {Xu},
  \citenamefont {Haule},\ and\ \citenamefont {Kotliar}}]{xu2013}%
  \BibitemOpen
  \bibfield  {author} {\bibinfo {author} {\bibfnamefont {W.}~\bibnamefont
  {Xu}}, \bibinfo {author} {\bibfnamefont {K.}~\bibnamefont {Haule}}, \ and\
  \bibinfo {author} {\bibfnamefont {G.}~\bibnamefont {Kotliar}},\ }\href
  {\doibase 10.1103/PhysRevLett.111.036401} {\bibfield  {journal} {\bibinfo
  {journal} {Physical Review Letters}\ }\textbf {\bibinfo {volume} {111}},\
  \bibinfo {pages} {036401} (\bibinfo {year} {2013})}\BibitemShut {NoStop}%
\bibitem [{\citenamefont {Haule}\ and\ \citenamefont
  {Kotliar}(2009)}]{haule2009-c}%
  \BibitemOpen
  \bibfield  {author} {\bibinfo {author} {\bibfnamefont {K.}~\bibnamefont
  {Haule}}\ and\ \bibinfo {author} {\bibfnamefont {G.}~\bibnamefont
  {Kotliar}},\ }\href {\doibase 10.1088/1367-2630/11/2/025021} {\bibfield
  {journal} {\bibinfo  {journal} {New Journal of Physics}\ }\textbf {\bibinfo
  {volume} {11}},\ \bibinfo {pages} {025021} (\bibinfo {year}
  {2009})}\BibitemShut {NoStop}%
\bibitem [{\citenamefont {Yin}\ \emph {et~al.}(2011)\citenamefont {Yin},
  \citenamefont {Haule},\ and\ \citenamefont {Kotliar}}]{yin2011-a}%
  \BibitemOpen
  \bibfield  {author} {\bibinfo {author} {\bibfnamefont {Z.~P.}\ \bibnamefont
  {Yin}}, \bibinfo {author} {\bibfnamefont {K.}~\bibnamefont {Haule}}, \ and\
  \bibinfo {author} {\bibfnamefont {G.}~\bibnamefont {Kotliar}},\ }\href
  {\doibase 10.1038/nphys1923} {\bibfield  {journal} {\bibinfo  {journal}
  {Nature Physics}\ }\textbf {\bibinfo {volume} {7}},\ \bibinfo {pages} {294}
  (\bibinfo {year} {2011})}\BibitemShut {NoStop}%
\bibitem [{\citenamefont {Mravlje}\ \emph {et~al.}(2011)\citenamefont
  {Mravlje}, \citenamefont {Aichhorn}, \citenamefont {Miyake}, \citenamefont
  {Haule}, \citenamefont {Kotliar},\ and\ \citenamefont
  {Georges}}]{mravlje2011}%
  \BibitemOpen
  \bibfield  {author} {\bibinfo {author} {\bibfnamefont {J.}~\bibnamefont
  {Mravlje}}, \bibinfo {author} {\bibfnamefont {M.}~\bibnamefont {Aichhorn}},
  \bibinfo {author} {\bibfnamefont {T.}~\bibnamefont {Miyake}}, \bibinfo
  {author} {\bibfnamefont {K.}~\bibnamefont {Haule}}, \bibinfo {author}
  {\bibfnamefont {G.}~\bibnamefont {Kotliar}}, \ and\ \bibinfo {author}
  {\bibfnamefont {A.}~\bibnamefont {Georges}},\ }\href {\doibase
  10.1103/PhysRevLett.106.096401} {\bibfield  {journal} {\bibinfo  {journal}
  {Physical Review Letters}\ }\textbf {\bibinfo {volume} {106}},\ \bibinfo
  {pages} {096401} (\bibinfo {year} {2011})}\BibitemShut {NoStop}%
\bibitem [{\citenamefont {Yin}\ \emph {et~al.}(2012)\citenamefont {Yin},
  \citenamefont {Haule},\ and\ \citenamefont {Kotliar}}]{yin2012}%
  \BibitemOpen
  \bibfield  {author} {\bibinfo {author} {\bibfnamefont {Z.~P.}\ \bibnamefont
  {Yin}}, \bibinfo {author} {\bibfnamefont {K.}~\bibnamefont {Haule}}, \ and\
  \bibinfo {author} {\bibfnamefont {G.}~\bibnamefont {Kotliar}},\ }\href
  {\doibase 10.1103/PhysRevB.86.195141} {\bibfield  {journal} {\bibinfo
  {journal} {Physical Review B}\ }\textbf {\bibinfo {volume} {86}},\ \bibinfo
  {pages} {195141} (\bibinfo {year} {2012})}\BibitemShut {NoStop}%
\bibitem [{\citenamefont {Georges}\ \emph {et~al.}(2013)\citenamefont
  {Georges}, \citenamefont {Medici},\ and\ \citenamefont
  {Mravlje}}]{georges2013}%
  \BibitemOpen
  \bibfield  {author} {\bibinfo {author} {\bibfnamefont {A.}~\bibnamefont
  {Georges}}, \bibinfo {author} {\bibfnamefont {L.~d.}\ \bibnamefont {Medici}},
  \ and\ \bibinfo {author} {\bibfnamefont {J.}~\bibnamefont {Mravlje}},\ }\href
  {\doibase 10.1146/annurev-conmatphys-020911-125045} {\bibfield  {journal}
  {\bibinfo  {journal} {Annual Review of Condensed Matter Physics}\ }\textbf
  {\bibinfo {volume} {4}},\ \bibinfo {pages} {137} (\bibinfo {year}
  {2013})}\BibitemShut {NoStop}%
\bibitem [{\citenamefont {Allen}\ \emph {et~al.}(1996)\citenamefont {Allen},
  \citenamefont {Berger}, \citenamefont {Chauvet}, \citenamefont {Forro},
  \citenamefont {Jarlborg}, \citenamefont {Junod}, \citenamefont {Revaz},\ and\
  \citenamefont {Santi}}]{allen1996}%
  \BibitemOpen
  \bibfield  {author} {\bibinfo {author} {\bibfnamefont {P.~B.}\ \bibnamefont
  {Allen}}, \bibinfo {author} {\bibfnamefont {H.}~\bibnamefont {Berger}},
  \bibinfo {author} {\bibfnamefont {O.}~\bibnamefont {Chauvet}}, \bibinfo
  {author} {\bibfnamefont {L.}~\bibnamefont {Forro}}, \bibinfo {author}
  {\bibfnamefont {T.}~\bibnamefont {Jarlborg}}, \bibinfo {author}
  {\bibfnamefont {A.}~\bibnamefont {Junod}}, \bibinfo {author} {\bibfnamefont
  {B.}~\bibnamefont {Revaz}}, \ and\ \bibinfo {author} {\bibfnamefont
  {G.}~\bibnamefont {Santi}},\ }\href {\doibase 10.1103/PhysRevB.53.4393}
  {\bibfield  {journal} {\bibinfo  {journal} {Physical Review B}\ }\textbf
  {\bibinfo {volume} {53}},\ \bibinfo {pages} {4393} (\bibinfo {year}
  {1996})}\BibitemShut {NoStop}%
\bibitem [{\citenamefont {Cao}\ \emph {et~al.}(1997)\citenamefont {Cao},
  \citenamefont {{McCall}}, \citenamefont {Shepard}, \citenamefont {Crow},\
  and\ \citenamefont {Guertin}}]{cao1997}%
  \BibitemOpen
  \bibfield  {author} {\bibinfo {author} {\bibfnamefont {G.}~\bibnamefont
  {Cao}}, \bibinfo {author} {\bibfnamefont {S.}~\bibnamefont {{McCall}}},
  \bibinfo {author} {\bibfnamefont {M.}~\bibnamefont {Shepard}}, \bibinfo
  {author} {\bibfnamefont {J.~E.}\ \bibnamefont {Crow}}, \ and\ \bibinfo
  {author} {\bibfnamefont {R.~P.}\ \bibnamefont {Guertin}},\ }\href {\doibase
  10.1103/PhysRevB.56.321} {\bibfield  {journal} {\bibinfo  {journal} {Physical
  Review B}\ }\textbf {\bibinfo {volume} {56}},\ \bibinfo {pages} {321}
  (\bibinfo {year} {1997})}\BibitemShut {NoStop}%
\bibitem [{\citenamefont {Shepard}\ \emph {et~al.}(1997)\citenamefont
  {Shepard}, \citenamefont {{McCall}}, \citenamefont {Cao},\ and\ \citenamefont
  {Crow}}]{shepard1997}%
  \BibitemOpen
  \bibfield  {author} {\bibinfo {author} {\bibfnamefont {M.}~\bibnamefont
  {Shepard}}, \bibinfo {author} {\bibfnamefont {S.}~\bibnamefont {{McCall}}},
  \bibinfo {author} {\bibfnamefont {G.}~\bibnamefont {Cao}}, \ and\ \bibinfo
  {author} {\bibfnamefont {J.~E.}\ \bibnamefont {Crow}},\ }\href {\doibase
  10.1063/1.365018} {\bibfield  {journal} {\bibinfo  {journal} {Journal of
  Applied Physics}\ }\textbf {\bibinfo {volume} {81}},\ \bibinfo {pages} {4978}
  (\bibinfo {year} {1997})}\BibitemShut {NoStop}%
\bibitem [{\citenamefont {Maeno}\ \emph {et~al.}(1997)\citenamefont {Maeno},
  \citenamefont {Yoshida}, \citenamefont {Hashimoto}, \citenamefont
  {Nishizaki}, \citenamefont {Ikeda}, \citenamefont {Nohara}, \citenamefont
  {Fujita}, \citenamefont {Mackenzie}, \citenamefont {Hussey}, \citenamefont
  {Bednorz},\ and\ \citenamefont {Lichtenberg}}]{maeno1997}%
  \BibitemOpen
  \bibfield  {author} {\bibinfo {author} {\bibfnamefont {Y.}~\bibnamefont
  {Maeno}}, \bibinfo {author} {\bibfnamefont {K.}~\bibnamefont {Yoshida}},
  \bibinfo {author} {\bibfnamefont {H.}~\bibnamefont {Hashimoto}}, \bibinfo
  {author} {\bibfnamefont {S.}~\bibnamefont {Nishizaki}}, \bibinfo {author}
  {\bibfnamefont {S.-i.}\ \bibnamefont {Ikeda}}, \bibinfo {author}
  {\bibfnamefont {M.}~\bibnamefont {Nohara}}, \bibinfo {author} {\bibfnamefont
  {T.}~\bibnamefont {Fujita}}, \bibinfo {author} {\bibfnamefont {A.~P.}\
  \bibnamefont {Mackenzie}}, \bibinfo {author} {\bibfnamefont {N.~E.}\
  \bibnamefont {Hussey}}, \bibinfo {author} {\bibfnamefont {J.~G.}\
  \bibnamefont {Bednorz}}, \ and\ \bibinfo {author} {\bibfnamefont
  {F.}~\bibnamefont {Lichtenberg}},\ }\href {\doibase 10.1143/JPSJ.66.1405}
  {\bibfield  {journal} {\bibinfo  {journal} {Journal of the Physical Society
  of Japan}\ }\textbf {\bibinfo {volume} {66}},\ \bibinfo {pages} {1405}
  (\bibinfo {year} {1997})}\BibitemShut {NoStop}%
\bibitem [{\citenamefont {Ikeda}\ \emph {et~al.}(2000)\citenamefont {Ikeda},
  \citenamefont {Maeno}, \citenamefont {Nakatsuji}, \citenamefont {Kosaka},\
  and\ \citenamefont {Uwatoko}}]{ikeda2000}%
  \BibitemOpen
  \bibfield  {author} {\bibinfo {author} {\bibfnamefont {S.}~\bibnamefont
  {Ikeda}}, \bibinfo {author} {\bibfnamefont {Y.}~\bibnamefont {Maeno}},
  \bibinfo {author} {\bibfnamefont {S.}~\bibnamefont {Nakatsuji}}, \bibinfo
  {author} {\bibfnamefont {M.}~\bibnamefont {Kosaka}}, \ and\ \bibinfo {author}
  {\bibfnamefont {Y.}~\bibnamefont {Uwatoko}},\ }\href {\doibase
  10.1103/PhysRevB.62.R6089} {\bibfield  {journal} {\bibinfo  {journal}
  {Physical Review B}\ }\textbf {\bibinfo {volume} {62}},\ \bibinfo {pages}
  {R6089} (\bibinfo {year} {2000})}\BibitemShut {NoStop}%
\bibitem [{\citenamefont {Bergemann}\ \emph {et~al.}(2003)\citenamefont
  {Bergemann}, \citenamefont {Mackenzie}, \citenamefont {Julian}, \citenamefont
  {Forsythe},\ and\ \citenamefont {Ohmichi}}]{bergemann2003}%
  \BibitemOpen
  \bibfield  {author} {\bibinfo {author} {\bibfnamefont {C.}~\bibnamefont
  {Bergemann}}, \bibinfo {author} {\bibfnamefont {A.~P.}\ \bibnamefont
  {Mackenzie}}, \bibinfo {author} {\bibfnamefont {S.~R.}\ \bibnamefont
  {Julian}}, \bibinfo {author} {\bibfnamefont {D.}~\bibnamefont {Forsythe}}, \
  and\ \bibinfo {author} {\bibfnamefont {E.}~\bibnamefont {Ohmichi}},\ }\href
  {\doibase 10.1080/00018730310001621737} {\bibfield  {journal} {\bibinfo
  {journal} {Advances in Physics}\ }\textbf {\bibinfo {volume} {52}},\ \bibinfo
  {pages} {639} (\bibinfo {year} {2003})}\BibitemShut {NoStop}%
\bibitem [{\citenamefont {Alexander}\ \emph {et~al.}(2005)\citenamefont
  {Alexander}, \citenamefont {{McCall}}, \citenamefont {Schlottmann},
  \citenamefont {Crow},\ and\ \citenamefont {Cao}}]{alexander2005}%
  \BibitemOpen
  \bibfield  {author} {\bibinfo {author} {\bibfnamefont {C.~S.}\ \bibnamefont
  {Alexander}}, \bibinfo {author} {\bibfnamefont {S.}~\bibnamefont {{McCall}}},
  \bibinfo {author} {\bibfnamefont {P.}~\bibnamefont {Schlottmann}}, \bibinfo
  {author} {\bibfnamefont {J.~E.}\ \bibnamefont {Crow}}, \ and\ \bibinfo
  {author} {\bibfnamefont {G.}~\bibnamefont {Cao}},\ }\href {\doibase
  10.1103/PhysRevB.72.024415} {\bibfield  {journal} {\bibinfo  {journal}
  {Physical Review B}\ }\textbf {\bibinfo {volume} {72}},\ \bibinfo {pages}
  {024415} (\bibinfo {year} {2005})}\BibitemShut {NoStop}%
\bibitem [{\citenamefont {Tamai}\ \emph {et~al.}(2008)\citenamefont {Tamai},
  \citenamefont {Allan}, \citenamefont {Mercure}, \citenamefont {Meevasana},
  \citenamefont {Dunkel}, \citenamefont {Lu}, \citenamefont {Perry},
  \citenamefont {Mackenzie}, \citenamefont {Singh}, \citenamefont {Shen},\ and\
  \citenamefont {Baumberger}}]{tamai2008}%
  \BibitemOpen
  \bibfield  {author} {\bibinfo {author} {\bibfnamefont {A.}~\bibnamefont
  {Tamai}}, \bibinfo {author} {\bibfnamefont {M.~P.}\ \bibnamefont {Allan}},
  \bibinfo {author} {\bibfnamefont {J.~F.}\ \bibnamefont {Mercure}}, \bibinfo
  {author} {\bibfnamefont {W.}~\bibnamefont {Meevasana}}, \bibinfo {author}
  {\bibfnamefont {R.}~\bibnamefont {Dunkel}}, \bibinfo {author} {\bibfnamefont
  {D.~H.}\ \bibnamefont {Lu}}, \bibinfo {author} {\bibfnamefont {R.~S.}\
  \bibnamefont {Perry}}, \bibinfo {author} {\bibfnamefont {A.~P.}\ \bibnamefont
  {Mackenzie}}, \bibinfo {author} {\bibfnamefont {D.~J.}\ \bibnamefont
  {Singh}}, \bibinfo {author} {\bibfnamefont {Z.}~\bibnamefont {Shen}}, \ and\
  \bibinfo {author} {\bibfnamefont {F.}~\bibnamefont {Baumberger}},\ }\href
  {\doibase 10.1103/PhysRevLett.101.026407} {\bibfield  {journal} {\bibinfo
  {journal} {Physical Review Letters}\ }\textbf {\bibinfo {volume} {101}},\
  \bibinfo {pages} {026407} (\bibinfo {year} {2008})}\BibitemShut {NoStop}%
\bibitem [{\citenamefont {Mercure}\ \emph {et~al.}(2009)\citenamefont
  {Mercure}, \citenamefont {Goh}, \citenamefont {{O{\textquoteright}Farrell}},
  \citenamefont {Perry}, \citenamefont {Sutherland}, \citenamefont {Rost},
  \citenamefont {Grigera}, \citenamefont {Borzi}, \citenamefont {Gegenwart},\
  and\ \citenamefont {Mackenzie}}]{mercure2009}%
  \BibitemOpen
  \bibfield  {author} {\bibinfo {author} {\bibfnamefont {J.}~\bibnamefont
  {Mercure}}, \bibinfo {author} {\bibfnamefont {S.~K.}\ \bibnamefont {Goh}},
  \bibinfo {author} {\bibfnamefont {E.~C.~T.}\ \bibnamefont
  {{O{\textquoteright}Farrell}}}, \bibinfo {author} {\bibfnamefont {R.~S.}\
  \bibnamefont {Perry}}, \bibinfo {author} {\bibfnamefont {M.~L.}\ \bibnamefont
  {Sutherland}}, \bibinfo {author} {\bibfnamefont {A.~W.}\ \bibnamefont
  {Rost}}, \bibinfo {author} {\bibfnamefont {S.~A.}\ \bibnamefont {Grigera}},
  \bibinfo {author} {\bibfnamefont {R.~A.}\ \bibnamefont {Borzi}}, \bibinfo
  {author} {\bibfnamefont {P.}~\bibnamefont {Gegenwart}}, \ and\ \bibinfo
  {author} {\bibfnamefont {A.~P.}\ \bibnamefont {Mackenzie}},\ }\href {\doibase
  10.1103/PhysRevLett.103.176401} {\bibfield  {journal} {\bibinfo  {journal}
  {Physical Review Letters}\ }\textbf {\bibinfo {volume} {103}},\ \bibinfo
  {pages} {176401} (\bibinfo {year} {2009})}\BibitemShut {NoStop}%
\bibitem [{\citenamefont {Iwasawa}\ \emph {et~al.}(2012)\citenamefont
  {Iwasawa}, \citenamefont {Yoshida}, \citenamefont {Hase}, \citenamefont
  {Shimada}, \citenamefont {Namatame}, \citenamefont {Taniguchi},\ and\
  \citenamefont {Aiura}}]{iwasawa2012}%
  \BibitemOpen
  \bibfield  {author} {\bibinfo {author} {\bibfnamefont {H.}~\bibnamefont
  {Iwasawa}}, \bibinfo {author} {\bibfnamefont {Y.}~\bibnamefont {Yoshida}},
  \bibinfo {author} {\bibfnamefont {I.}~\bibnamefont {Hase}}, \bibinfo {author}
  {\bibfnamefont {K.}~\bibnamefont {Shimada}}, \bibinfo {author} {\bibfnamefont
  {H.}~\bibnamefont {Namatame}}, \bibinfo {author} {\bibfnamefont
  {M.}~\bibnamefont {Taniguchi}}, \ and\ \bibinfo {author} {\bibfnamefont
  {Y.}~\bibnamefont {Aiura}},\ }\href {\doibase 10.1103/PhysRevLett.109.066404}
  {\bibfield  {journal} {\bibinfo  {journal} {Physical Review Letters}\
  }\textbf {\bibinfo {volume} {109}},\ \bibinfo {pages} {066404} (\bibinfo
  {year} {2012})}\BibitemShut {NoStop}%
\bibitem [{\citenamefont {Shai}\ \emph {et~al.}(2013)\citenamefont {Shai},
  \citenamefont {Adamo}, \citenamefont {Shen}, \citenamefont {Brooks},
  \citenamefont {Harter}, \citenamefont {Monkman}, \citenamefont {Burganov},
  \citenamefont {Schlom},\ and\ \citenamefont {Shen}}]{shai2013}%
  \BibitemOpen
  \bibfield  {author} {\bibinfo {author} {\bibfnamefont {D.~E.}\ \bibnamefont
  {Shai}}, \bibinfo {author} {\bibfnamefont {C.}~\bibnamefont {Adamo}},
  \bibinfo {author} {\bibfnamefont {D.~W.}\ \bibnamefont {Shen}}, \bibinfo
  {author} {\bibfnamefont {C.~M.}\ \bibnamefont {Brooks}}, \bibinfo {author}
  {\bibfnamefont {J.~W.}\ \bibnamefont {Harter}}, \bibinfo {author}
  {\bibfnamefont {E.~J.}\ \bibnamefont {Monkman}}, \bibinfo {author}
  {\bibfnamefont {B.}~\bibnamefont {Burganov}}, \bibinfo {author}
  {\bibfnamefont {D.~G.}\ \bibnamefont {Schlom}}, \ and\ \bibinfo {author}
  {\bibfnamefont {K.~M.}\ \bibnamefont {Shen}},\ }\href {\doibase
  10.1103/PhysRevLett.110.087004} {\bibfield  {journal} {\bibinfo  {journal}
  {Physical Review Letters}\ }\textbf {\bibinfo {volume} {110}},\ \bibinfo
  {pages} {087004} (\bibinfo {year} {2013})}\BibitemShut {NoStop}%
\bibitem [{\citenamefont {Veenstra}\ \emph {et~al.}(2013)\citenamefont
  {Veenstra}, \citenamefont {Zhu}, \citenamefont {Ludbrook}, \citenamefont
  {Capsoni}, \citenamefont {Levy}, \citenamefont {Nicolaou}, \citenamefont
  {Rosen}, \citenamefont {Comin}, \citenamefont {Kittaka}, \citenamefont
  {Maeno}, \citenamefont {Elfimov},\ and\ \citenamefont
  {Damascelli}}]{veenstra2013}%
  \BibitemOpen
  \bibfield  {author} {\bibinfo {author} {\bibfnamefont {C.~N.}\ \bibnamefont
  {Veenstra}}, \bibinfo {author} {\bibfnamefont {Z.}~\bibnamefont {Zhu}},
  \bibinfo {author} {\bibfnamefont {B.}~\bibnamefont {Ludbrook}}, \bibinfo
  {author} {\bibfnamefont {M.}~\bibnamefont {Capsoni}}, \bibinfo {author}
  {\bibfnamefont {G.}~\bibnamefont {Levy}}, \bibinfo {author} {\bibfnamefont
  {A.}~\bibnamefont {Nicolaou}}, \bibinfo {author} {\bibfnamefont {J.~A.}\
  \bibnamefont {Rosen}}, \bibinfo {author} {\bibfnamefont {R.}~\bibnamefont
  {Comin}}, \bibinfo {author} {\bibfnamefont {S.}~\bibnamefont {Kittaka}},
  \bibinfo {author} {\bibfnamefont {Y.}~\bibnamefont {Maeno}}, \bibinfo
  {author} {\bibfnamefont {I.~S.}\ \bibnamefont {Elfimov}}, \ and\ \bibinfo
  {author} {\bibfnamefont {A.}~\bibnamefont {Damascelli}},\ }\href {\doibase
  10.1103/PhysRevLett.110.097004} {\bibfield  {journal} {\bibinfo  {journal}
  {Physical Review Letters}\ }\textbf {\bibinfo {volume} {110}},\ \bibinfo
  {pages} {097004} (\bibinfo {year} {2013})}\BibitemShut {NoStop}%
\bibitem [{\citenamefont {Allan}\ \emph {et~al.}(2013)\citenamefont {Allan},
  \citenamefont {Tamai}, \citenamefont {Rozbicki}, \citenamefont {Fischer},
  \citenamefont {Voss}, \citenamefont {King}, \citenamefont {Meevasana},
  \citenamefont {Thirupathaiah}, \citenamefont {Rienks}, \citenamefont {Fink},
  \citenamefont {Tennant}, \citenamefont {Perry}, \citenamefont {Mercure},
  \citenamefont {Wang}, \citenamefont {Lee}, \citenamefont {Fennie},
  \citenamefont {Kim}, \citenamefont {Lawler}, \citenamefont {Shen},
  \citenamefont {Mackenzie}, \citenamefont {Shen},\ and\ \citenamefont
  {Baumberger}}]{allan2013}%
  \BibitemOpen
  \bibfield  {author} {\bibinfo {author} {\bibfnamefont {M.~P.}\ \bibnamefont
  {Allan}}, \bibinfo {author} {\bibfnamefont {A.}~\bibnamefont {Tamai}},
  \bibinfo {author} {\bibfnamefont {E.}~\bibnamefont {Rozbicki}}, \bibinfo
  {author} {\bibfnamefont {M.~H.}\ \bibnamefont {Fischer}}, \bibinfo {author}
  {\bibfnamefont {J.}~\bibnamefont {Voss}}, \bibinfo {author} {\bibfnamefont
  {P.~D.~C.}\ \bibnamefont {King}}, \bibinfo {author} {\bibfnamefont
  {W.}~\bibnamefont {Meevasana}}, \bibinfo {author} {\bibfnamefont
  {S.}~\bibnamefont {Thirupathaiah}}, \bibinfo {author} {\bibfnamefont
  {E.}~\bibnamefont {Rienks}}, \bibinfo {author} {\bibfnamefont
  {J.}~\bibnamefont {Fink}}, \bibinfo {author} {\bibfnamefont {D.~A.}\
  \bibnamefont {Tennant}}, \bibinfo {author} {\bibfnamefont {R.~S.}\
  \bibnamefont {Perry}}, \bibinfo {author} {\bibfnamefont {J.~F.}\ \bibnamefont
  {Mercure}}, \bibinfo {author} {\bibfnamefont {M.~A.}\ \bibnamefont {Wang}},
  \bibinfo {author} {\bibfnamefont {J.}~\bibnamefont {Lee}}, \bibinfo {author}
  {\bibfnamefont {C.~J.}\ \bibnamefont {Fennie}}, \bibinfo {author}
  {\bibfnamefont {E.}~\bibnamefont {Kim}}, \bibinfo {author} {\bibfnamefont
  {M.~J.}\ \bibnamefont {Lawler}}, \bibinfo {author} {\bibfnamefont {K.~M.}\
  \bibnamefont {Shen}}, \bibinfo {author} {\bibfnamefont {A.~P.}\ \bibnamefont
  {Mackenzie}}, \bibinfo {author} {\bibfnamefont {Z.}~\bibnamefont {Shen}}, \
  and\ \bibinfo {author} {\bibfnamefont {F.}~\bibnamefont {Baumberger}},\
  }\href {\doibase 10.1088/1367-2630/15/6/063029} {\bibfield  {journal}
  {\bibinfo  {journal} {New Journal of Physics}\ }\textbf {\bibinfo {volume}
  {15}},\ \bibinfo {pages} {063029} (\bibinfo {year} {2013})}\BibitemShut
  {NoStop}%
\bibitem [{\citenamefont {Schneider}\ \emph {et~al.}(2014)\citenamefont
  {Schneider}, \citenamefont {Geiger}, \citenamefont {Esser}, \citenamefont
  {Pracht}, \citenamefont {Stingl}, \citenamefont {Tokiwa}, \citenamefont
  {Moshnyaga}, \citenamefont {Sheikin}, \citenamefont {Mravlje}, \citenamefont
  {Scheffler},\ and\ \citenamefont {Gegenwart}}]{schneider2014}%
  \BibitemOpen
  \bibfield  {author} {\bibinfo {author} {\bibfnamefont {M.}~\bibnamefont
  {Schneider}}, \bibinfo {author} {\bibfnamefont {D.}~\bibnamefont {Geiger}},
  \bibinfo {author} {\bibfnamefont {S.}~\bibnamefont {Esser}}, \bibinfo
  {author} {\bibfnamefont {U.~S.}\ \bibnamefont {Pracht}}, \bibinfo {author}
  {\bibfnamefont {C.}~\bibnamefont {Stingl}}, \bibinfo {author} {\bibfnamefont
  {Y.}~\bibnamefont {Tokiwa}}, \bibinfo {author} {\bibfnamefont
  {V.}~\bibnamefont {Moshnyaga}}, \bibinfo {author} {\bibfnamefont
  {I.}~\bibnamefont {Sheikin}}, \bibinfo {author} {\bibfnamefont
  {J.}~\bibnamefont {Mravlje}}, \bibinfo {author} {\bibfnamefont
  {M.}~\bibnamefont {Scheffler}}, \ and\ \bibinfo {author} {\bibfnamefont
  {P.}~\bibnamefont {Gegenwart}},\ }\href {\doibase
  10.1103/PhysRevLett.112.206403} {\bibfield  {journal} {\bibinfo  {journal}
  {Physical Review Letters}\ }\textbf {\bibinfo {volume} {112}},\ \bibinfo
  {pages} {206403} (\bibinfo {year} {2014})}\BibitemShut {NoStop}%
\bibitem [{\citenamefont {Cao}\ \emph {et~al.}(2004)\citenamefont {Cao},
  \citenamefont {Song}, \citenamefont {Sun},\ and\ \citenamefont
  {Lin}}]{cao2004}%
  \BibitemOpen
  \bibfield  {author} {\bibinfo {author} {\bibfnamefont {G.}~\bibnamefont
  {Cao}}, \bibinfo {author} {\bibfnamefont {W.}~\bibnamefont {Song}}, \bibinfo
  {author} {\bibfnamefont {Y.}~\bibnamefont {Sun}}, \ and\ \bibinfo {author}
  {\bibfnamefont {X.}~\bibnamefont {Lin}},\ }\href {\doibase
  10.1016/j.ssc.2004.03.001} {\bibfield  {journal} {\bibinfo  {journal} {Solid
  State Communications}\ }\textbf {\bibinfo {volume} {131}},\ \bibinfo {pages}
  {331} (\bibinfo {year} {2004})}\BibitemShut {NoStop}%
\bibitem [{\citenamefont {Hussey}\ \emph {et~al.}(1998)\citenamefont {Hussey},
  \citenamefont {Mackenzie}, \citenamefont {Cooper}, \citenamefont {Maeno},
  \citenamefont {Nishizaki},\ and\ \citenamefont {Fujita}}]{hussey1998}%
  \BibitemOpen
  \bibfield  {author} {\bibinfo {author} {\bibfnamefont {N.~E.}\ \bibnamefont
  {Hussey}}, \bibinfo {author} {\bibfnamefont {A.~P.}\ \bibnamefont
  {Mackenzie}}, \bibinfo {author} {\bibfnamefont {J.~R.}\ \bibnamefont
  {Cooper}}, \bibinfo {author} {\bibfnamefont {Y.}~\bibnamefont {Maeno}},
  \bibinfo {author} {\bibfnamefont {S.}~\bibnamefont {Nishizaki}}, \ and\
  \bibinfo {author} {\bibfnamefont {T.}~\bibnamefont {Fujita}},\ }\href
  {\doibase 10.1103/PhysRevB.57.5505} {\bibfield  {journal} {\bibinfo
  {journal} {Physical Review B}\ }\textbf {\bibinfo {volume} {57}},\ \bibinfo
  {pages} {5505} (\bibinfo {year} {1998})}\BibitemShut {NoStop}%
\bibitem [{\citenamefont {Tyler}\ \emph {et~al.}(1998)\citenamefont {Tyler},
  \citenamefont {Mackenzie}, \citenamefont {{NishiZaki}},\ and\ \citenamefont
  {Maeno}}]{tyler1998}%
  \BibitemOpen
  \bibfield  {author} {\bibinfo {author} {\bibfnamefont {A.~W.}\ \bibnamefont
  {Tyler}}, \bibinfo {author} {\bibfnamefont {A.~P.}\ \bibnamefont
  {Mackenzie}}, \bibinfo {author} {\bibfnamefont {S.}~\bibnamefont
  {{NishiZaki}}}, \ and\ \bibinfo {author} {\bibfnamefont {Y.}~\bibnamefont
  {Maeno}},\ }\href {\doibase 10.1103/PhysRevB.58.R10107} {\bibfield  {journal}
  {\bibinfo  {journal} {Physical Review B}\ }\textbf {\bibinfo {volume} {58}},\
  \bibinfo {pages} {R10107} (\bibinfo {year} {1998})}\BibitemShut {NoStop}%
\bibitem [{\citenamefont {Bruin}\ \emph {et~al.}(2013)\citenamefont {Bruin},
  \citenamefont {Sakai}, \citenamefont {Perry},\ and\ \citenamefont
  {Mackenzie}}]{bruin2013}%
  \BibitemOpen
  \bibfield  {author} {\bibinfo {author} {\bibfnamefont {J.~a.~N.}\
  \bibnamefont {Bruin}}, \bibinfo {author} {\bibfnamefont {H.}~\bibnamefont
  {Sakai}}, \bibinfo {author} {\bibfnamefont {R.~S.}\ \bibnamefont {Perry}}, \
  and\ \bibinfo {author} {\bibfnamefont {A.~P.}\ \bibnamefont {Mackenzie}},\
  }\href {\doibase 10.1126/science.1227612} {\bibfield  {journal} {\bibinfo
  {journal} {Science}\ }\textbf {\bibinfo {volume} {339}},\ \bibinfo {pages}
  {804} (\bibinfo {year} {2013})},\ \bibinfo {note} {{PMID:}
  23413351}\BibitemShut {NoStop}%
\bibitem [{\citenamefont {Klein}\ \emph {et~al.}(1996)\citenamefont {Klein},
  \citenamefont {Dodge}, \citenamefont {Ahn}, \citenamefont {Snyder},
  \citenamefont {Geballe}, \citenamefont {Beasley},\ and\ \citenamefont
  {Kapitulnik}}]{klein1996}%
  \BibitemOpen
  \bibfield  {author} {\bibinfo {author} {\bibfnamefont {L.}~\bibnamefont
  {Klein}}, \bibinfo {author} {\bibfnamefont {J.~S.}\ \bibnamefont {Dodge}},
  \bibinfo {author} {\bibfnamefont {C.~H.}\ \bibnamefont {Ahn}}, \bibinfo
  {author} {\bibfnamefont {G.~J.}\ \bibnamefont {Snyder}}, \bibinfo {author}
  {\bibfnamefont {T.~H.}\ \bibnamefont {Geballe}}, \bibinfo {author}
  {\bibfnamefont {M.~R.}\ \bibnamefont {Beasley}}, \ and\ \bibinfo {author}
  {\bibfnamefont {A.}~\bibnamefont {Kapitulnik}},\ }\href {\doibase
  10.1103/PhysRevLett.77.2774} {\bibfield  {journal} {\bibinfo  {journal}
  {Physical Review Letters}\ }\textbf {\bibinfo {volume} {77}},\ \bibinfo
  {pages} {2774} (\bibinfo {year} {1996})}\BibitemShut {NoStop}%
\bibitem [{\citenamefont {Kotliar}\ \emph {et~al.}(2006)\citenamefont
  {Kotliar}, \citenamefont {Savrasov}, \citenamefont {Haule}, \citenamefont
  {Oudovenko}, \citenamefont {Parcollet},\ and\ \citenamefont
  {Marianetti}}]{kotliar2006}%
  \BibitemOpen
  \bibfield  {author} {\bibinfo {author} {\bibfnamefont {G.}~\bibnamefont
  {Kotliar}}, \bibinfo {author} {\bibfnamefont {S.}~\bibnamefont {Savrasov}},
  \bibinfo {author} {\bibfnamefont {K.}~\bibnamefont {Haule}}, \bibinfo
  {author} {\bibfnamefont {V.}~\bibnamefont {Oudovenko}}, \bibinfo {author}
  {\bibfnamefont {O.}~\bibnamefont {Parcollet}}, \ and\ \bibinfo {author}
  {\bibfnamefont {C.}~\bibnamefont {Marianetti}},\ }\href {\doibase
  10.1103/RevModPhys.78.865} {\bibfield  {journal} {\bibinfo  {journal}
  {Reviews of Modern Physics}\ }\textbf {\bibinfo {volume} {78}},\ \bibinfo
  {pages} {865} (\bibinfo {year} {2006})}\BibitemShut {NoStop}%
\bibitem [{\citenamefont {Jakobi}\ \emph {et~al.}(2011)\citenamefont {Jakobi},
  \citenamefont {Kanungo}, \citenamefont {Sarkar}, \citenamefont {Schmitt},\
  and\ \citenamefont {{Saha-Dasgupta}}}]{jakobi2011}%
  \BibitemOpen
  \bibfield  {author} {\bibinfo {author} {\bibfnamefont {E.}~\bibnamefont
  {Jakobi}}, \bibinfo {author} {\bibfnamefont {S.}~\bibnamefont {Kanungo}},
  \bibinfo {author} {\bibfnamefont {S.}~\bibnamefont {Sarkar}}, \bibinfo
  {author} {\bibfnamefont {S.}~\bibnamefont {Schmitt}}, \ and\ \bibinfo
  {author} {\bibfnamefont {T.}~\bibnamefont {{Saha-Dasgupta}}},\ }\href
  {\doibase 10.1103/PhysRevB.83.041103} {\bibfield  {journal} {\bibinfo
  {journal} {Physical Review B}\ }\textbf {\bibinfo {volume} {83}},\ \bibinfo
  {pages} {041103} (\bibinfo {year} {2011})}\BibitemShut {NoStop}%
\bibitem [{\citenamefont {Dang}\ \emph {et~al.}(2014)\citenamefont {Dang},
  \citenamefont {Mravlje}, \citenamefont {Georges},\ and\ \citenamefont
  {Millis}}]{dang2014}%
  \BibitemOpen
  \bibfield  {author} {\bibinfo {author} {\bibfnamefont {H.~T.}\ \bibnamefont
  {Dang}}, \bibinfo {author} {\bibfnamefont {J.}~\bibnamefont {Mravlje}},
  \bibinfo {author} {\bibfnamefont {A.}~\bibnamefont {Georges}}, \ and\
  \bibinfo {author} {\bibfnamefont {A.~J.}\ \bibnamefont {Millis}},\ }\href
  {http://arxiv.org/abs/1412.7803} {\bibfield  {journal} {\bibinfo  {journal}
  {{arXiv:1412.7803} [cond-mat]}\ } (\bibinfo {year} {2014})},\ \bibinfo {note}
  {{arXiv:} 1412.7803}\BibitemShut {NoStop}%
\bibitem [{\citenamefont {Dang}\ \emph {et~al.}(2015)\citenamefont {Dang},
  \citenamefont {Mravlje}, \citenamefont {Georges},\ and\ \citenamefont
  {Millis}}]{dang2015}%
  \BibitemOpen
  \bibfield  {author} {\bibinfo {author} {\bibfnamefont {H.~T.}\ \bibnamefont
  {Dang}}, \bibinfo {author} {\bibfnamefont {J.}~\bibnamefont {Mravlje}},
  \bibinfo {author} {\bibfnamefont {A.}~\bibnamefont {Georges}}, \ and\
  \bibinfo {author} {\bibfnamefont {A.~J.}\ \bibnamefont {Millis}},\ }\href
  {http://arxiv.org/abs/1501.03964} {\bibfield  {journal} {\bibinfo  {journal}
  {{arXiv:1501.03964} [cond-mat]}\ } (\bibinfo {year} {2015})},\ \bibinfo
  {note} {{arXiv:} 1501.03964}\BibitemShut {NoStop}%
\bibitem [{\citenamefont {Haule}\ \emph {et~al.}(2010)\citenamefont {Haule},
  \citenamefont {Yee},\ and\ \citenamefont {Kim}}]{haule2010}%
  \BibitemOpen
  \bibfield  {author} {\bibinfo {author} {\bibfnamefont {K.}~\bibnamefont
  {Haule}}, \bibinfo {author} {\bibfnamefont {C.}~\bibnamefont {Yee}}, \ and\
  \bibinfo {author} {\bibfnamefont {K.}~\bibnamefont {Kim}},\ }\href {\doibase
  10.1103/PhysRevB.81.195107} {\bibfield  {journal} {\bibinfo  {journal}
  {Physical Review B}\ }\textbf {\bibinfo {volume} {81}},\ \bibinfo {pages}
  {195107} (\bibinfo {year} {2010})}\BibitemShut {NoStop}%
\bibitem [{\citenamefont {Blaha}\ \emph {et~al.}(2001)\citenamefont {Blaha},
  \citenamefont {Schwarz}, \citenamefont {Madsen}, \citenamefont {Kvasnicka},\
  and\ \citenamefont {Luitz}}]{blaha2001}%
  \BibitemOpen
  \bibfield  {author} {\bibinfo {author} {\bibfnamefont {P.}~\bibnamefont
  {Blaha}}, \bibinfo {author} {\bibfnamefont {K.}~\bibnamefont {Schwarz}},
  \bibinfo {author} {\bibfnamefont {G.~K.~H.}\ \bibnamefont {Madsen}}, \bibinfo
  {author} {\bibfnamefont {D.}~\bibnamefont {Kvasnicka}}, \ and\ \bibinfo
  {author} {\bibfnamefont {J.}~\bibnamefont {Luitz}},\ }\href@noop {} {\emph
  {\bibinfo {title} {{WIEN2K}, An Augmented Plane Wave + Local Orbitals Program
  for Calculating Crystal Properties}}}\ (\bibinfo  {publisher} {Karlheinz
  Schwarz, Techn. Universit\"{a}t Wien, Austria},\ \bibinfo {address} {Wien,
  Austria},\ \bibinfo {year} {2001})\BibitemShut {NoStop}%
\bibitem [{\citenamefont {Haule}(2007)}]{haule2007}%
  \BibitemOpen
  \bibfield  {author} {\bibinfo {author} {\bibfnamefont {K.}~\bibnamefont
  {Haule}},\ }\href {\doibase 10.1103/PhysRevB.75.155113} {\bibfield  {journal}
  {\bibinfo  {journal} {Physical Review B}\ }\textbf {\bibinfo {volume} {75}},\
  \bibinfo {pages} {155113} (\bibinfo {year} {2007})}\BibitemShut {NoStop}%
\bibitem [{\citenamefont {Werner}\ \emph {et~al.}(2006)\citenamefont {Werner},
  \citenamefont {Comanac}, \citenamefont {de' Medici}, \citenamefont {Troyer},\
  and\ \citenamefont {Millis}}]{werner2006}%
  \BibitemOpen
  \bibfield  {author} {\bibinfo {author} {\bibfnamefont {P.}~\bibnamefont
  {Werner}}, \bibinfo {author} {\bibfnamefont {A.}~\bibnamefont {Comanac}},
  \bibinfo {author} {\bibfnamefont {L.}~\bibnamefont {de' Medici}}, \bibinfo
  {author} {\bibfnamefont {M.}~\bibnamefont {Troyer}}, \ and\ \bibinfo {author}
  {\bibfnamefont {A.~J.}\ \bibnamefont {Millis}},\ }\href {\doibase
  10.1103/PhysRevLett.97.076405} {\bibfield  {journal} {\bibinfo  {journal}
  {Physical Review Letters}\ }\textbf {\bibinfo {volume} {97}},\ \bibinfo
  {pages} {076405} (\bibinfo {year} {2006})}\BibitemShut {NoStop}%
\bibitem [{\citenamefont {Lee}\ \emph {et~al.}(2004)\citenamefont {Lee},
  \citenamefont {Lee}, \citenamefont {Noh}, \citenamefont {Nakatsuji},
  \citenamefont {Fukazawa}, \citenamefont {Perry}, \citenamefont {Maeno},
  \citenamefont {Yoshida}, \citenamefont {Ikeda}, \citenamefont {Yu},\ and\
  \citenamefont {Eom}}]{lee2004-a}%
  \BibitemOpen
  \bibfield  {author} {\bibinfo {author} {\bibfnamefont {J.~S.}\ \bibnamefont
  {Lee}}, \bibinfo {author} {\bibfnamefont {Y.~S.}\ \bibnamefont {Lee}},
  \bibinfo {author} {\bibfnamefont {T.~W.}\ \bibnamefont {Noh}}, \bibinfo
  {author} {\bibfnamefont {S.}~\bibnamefont {Nakatsuji}}, \bibinfo {author}
  {\bibfnamefont {H.}~\bibnamefont {Fukazawa}}, \bibinfo {author}
  {\bibfnamefont {R.~S.}\ \bibnamefont {Perry}}, \bibinfo {author}
  {\bibfnamefont {Y.}~\bibnamefont {Maeno}}, \bibinfo {author} {\bibfnamefont
  {Y.}~\bibnamefont {Yoshida}}, \bibinfo {author} {\bibfnamefont {S.~I.}\
  \bibnamefont {Ikeda}}, \bibinfo {author} {\bibfnamefont {J.}~\bibnamefont
  {Yu}}, \ and\ \bibinfo {author} {\bibfnamefont {C.~B.}\ \bibnamefont {Eom}},\
  }\href {\doibase 10.1103/PhysRevB.70.085103} {\bibfield  {journal} {\bibinfo
  {journal} {Physical Review B}\ }\textbf {\bibinfo {volume} {70}},\ \bibinfo
  {pages} {085103} (\bibinfo {year} {2004})}\BibitemShut {NoStop}%
\bibitem [{\citenamefont {Genish}\ \emph {et~al.}(2007)\citenamefont {Genish},
  \citenamefont {Klein}, \citenamefont {Reiner},\ and\ \citenamefont
  {Beasley}}]{genish2007}%
  \BibitemOpen
  \bibfield  {author} {\bibinfo {author} {\bibfnamefont {I.}~\bibnamefont
  {Genish}}, \bibinfo {author} {\bibfnamefont {L.}~\bibnamefont {Klein}},
  \bibinfo {author} {\bibfnamefont {J.~W.}\ \bibnamefont {Reiner}}, \ and\
  \bibinfo {author} {\bibfnamefont {M.~R.}\ \bibnamefont {Beasley}},\ }\href
  {\doibase 10.1103/PhysRevB.75.125108} {\bibfield  {journal} {\bibinfo
  {journal} {Physical Review B}\ }\textbf {\bibinfo {volume} {75}},\ \bibinfo
  {pages} {125108} (\bibinfo {year} {2007})}\BibitemShut {NoStop}%
\bibitem [{\citenamefont {Proffit}\ \emph {et~al.}(2008)\citenamefont
  {Proffit}, \citenamefont {Jang}, \citenamefont {Lee}, \citenamefont {Nelson},
  \citenamefont {Pan}, \citenamefont {Rzchowski},\ and\ \citenamefont
  {Eom}}]{proffit2008}%
  \BibitemOpen
  \bibfield  {author} {\bibinfo {author} {\bibfnamefont {D.~L.}\ \bibnamefont
  {Proffit}}, \bibinfo {author} {\bibfnamefont {H.~W.}\ \bibnamefont {Jang}},
  \bibinfo {author} {\bibfnamefont {S.}~\bibnamefont {Lee}}, \bibinfo {author}
  {\bibfnamefont {C.~T.}\ \bibnamefont {Nelson}}, \bibinfo {author}
  {\bibfnamefont {X.~Q.}\ \bibnamefont {Pan}}, \bibinfo {author} {\bibfnamefont
  {M.~S.}\ \bibnamefont {Rzchowski}}, \ and\ \bibinfo {author} {\bibfnamefont
  {C.~B.}\ \bibnamefont {Eom}},\ }\href {\doibase 10.1063/1.2979237} {\bibfield
   {journal} {\bibinfo  {journal} {Applied Physics Letters}\ }\textbf {\bibinfo
  {volume} {93}},\ \bibinfo {pages} {111912} (\bibinfo {year}
  {2008})}\BibitemShut {NoStop}%
\bibitem [{\citenamefont {Singh}(1995)}]{singh1995}%
  \BibitemOpen
  \bibfield  {author} {\bibinfo {author} {\bibfnamefont {D.~J.}\ \bibnamefont
  {Singh}},\ }\href {\doibase 10.1103/PhysRevB.52.1358} {\bibfield  {journal}
  {\bibinfo  {journal} {Physical Review B}\ }\textbf {\bibinfo {volume} {52}},\
  \bibinfo {pages} {1358} (\bibinfo {year} {1995})}\BibitemShut {NoStop}%
\bibitem [{\citenamefont {Deng}\ \emph {et~al.}(2014)\citenamefont {Deng},
  \citenamefont {Sternbach}, \citenamefont {Haule}, \citenamefont {Basov},\
  and\ \citenamefont {Kotliar}}]{deng2014}%
  \BibitemOpen
  \bibfield  {author} {\bibinfo {author} {\bibfnamefont {X.}~\bibnamefont
  {Deng}}, \bibinfo {author} {\bibfnamefont {A.}~\bibnamefont {Sternbach}},
  \bibinfo {author} {\bibfnamefont {K.}~\bibnamefont {Haule}}, \bibinfo
  {author} {\bibfnamefont {D.~N.}\ \bibnamefont {Basov}}, \ and\ \bibinfo
  {author} {\bibfnamefont {G.}~\bibnamefont {Kotliar}},\ }\href {\doibase
  10.1103/PhysRevLett.113.246404} {\bibfield  {journal} {\bibinfo  {journal}
  {Physical Review Letters}\ }\textbf {\bibinfo {volume} {113}},\ \bibinfo
  {pages} {246404} (\bibinfo {year} {2014})}\BibitemShut {NoStop}%
\bibitem [{\citenamefont {Kamal}\ \emph {et~al.}(2005)\citenamefont {Kamal},
  \citenamefont {Dodge}, \citenamefont {Kim},\ and\ \citenamefont
  {Eom}}]{kamal2005}%
  \BibitemOpen
  \bibfield  {author} {\bibinfo {author} {\bibfnamefont {S.}~\bibnamefont
  {Kamal}}, \bibinfo {author} {\bibfnamefont {J.~S.}\ \bibnamefont {Dodge}},
  \bibinfo {author} {\bibfnamefont {D.}~\bibnamefont {Kim}}, \ and\ \bibinfo
  {author} {\bibfnamefont {C.}~\bibnamefont {Eom}}\ }(\bibinfo  {publisher}
  {Optical Society of America},\ \bibinfo {year} {2005})\ p.\ \bibinfo {pages}
  {QWC2}\BibitemShut {NoStop}%
\bibitem [{sup()}]{supp}%
  \BibitemOpen
  \href@noop {} {}\bibinfo {note} {See Supplementary Materials at [url], for
  the Weiss function and the local Green's functions for all the orbitals, as
  well as the relation of the correlation strength in $d_{xy}$ orbital and the
  structure parameters}\BibitemShut {NoStop}%
\bibitem [{\citenamefont {Chang}\ \emph {et~al.}(2009)\citenamefont {Chang},
  \citenamefont {Kim}, \citenamefont {Phark}, \citenamefont {Kim},
  \citenamefont {Yu},\ and\ \citenamefont {Noh}}]{chang2009}%
  \BibitemOpen
  \bibfield  {author} {\bibinfo {author} {\bibfnamefont {Y.~J.}\ \bibnamefont
  {Chang}}, \bibinfo {author} {\bibfnamefont {C.~H.}\ \bibnamefont {Kim}},
  \bibinfo {author} {\bibfnamefont {S.}~\bibnamefont {Phark}}, \bibinfo
  {author} {\bibfnamefont {Y.~S.}\ \bibnamefont {Kim}}, \bibinfo {author}
  {\bibfnamefont {J.}~\bibnamefont {Yu}}, \ and\ \bibinfo {author}
  {\bibfnamefont {T.~W.}\ \bibnamefont {Noh}},\ }\href {\doibase
  10.1103/PhysRevLett.103.057201} {\bibfield  {journal} {\bibinfo  {journal}
  {Physical Review Letters}\ }\textbf {\bibinfo {volume} {103}},\ \bibinfo
  {pages} {057201} (\bibinfo {year} {2009})}\BibitemShut {NoStop}%
\end{thebibliography}
\end{document}